\documentclass[twoside]{article}
\usepackage{fleqn,espcrc2}
  
\usepackage{epsfig}
\usepackage[figuresright]{rotating}

\def\aprle{\buildrel < \over {_{\sim}}} 
\def\aprge{\buildrel > \over {_{\sim}}} 

\title{Cosmic Ray Astrophysics  and Hadronic  Interactions}

\author{Paolo Lipari
       \address{INFN sez. Roma 1, and 
       Dipartimento di Fisica, Universit\`a di Roma, ``La Sapienza''.}}
\begin{document}

\begin{abstract}
Research in cosmic rays is  now nearly a century old, but most of the
fundamental questions in this  field remain unanswered,
on the other hand the perspectives  of future studies   in the next decade
are very bright. New detectors  will provide higher  quality data in
the entire energy range  from $10^8$ to $10^{20}$~eV (or more if
particles of higher  energy have non negligible fluxes), moreover  cosmic ray
astrophysics  must now  be considered, together with  gamma, neutrino and
gravitational wave astronomy, as one of the subfields
of high energy astrophysics,
and  using information from these  four  ``messengers''   there is the
potential of a detailed understanding of the origin of the
 high energy radiation in the universe. 
High energy cosmic  rays  are measured indirectly  observing the 
showers  they generate in the atmosphere, and a 
correct  and detailed  interpretation of  these  measurements 
will  require  an improved understanding of the properties of 
hadronic interactions.  The new collider  experiments, and in particular 
the LHC project at CERN  offer the unique possibility to 
perform measurements  of great value for 
cosmic  ray  astrophysics.
 It is of  great 
importance for   cosmic research that this possibility is fully exploited 
with the appropriate instrumentation and analysis.
\end{abstract}

\maketitle

%\section{Introduction}
\section{High Energy Astrophysics \\
and Fundamental Science}
Progress in fundamental  science  requires the study  of
``extreme  physical systems'', where the deeper structure of the
physical  laws  can become  visible,
such ``extreme systems'' can  be constructed in the laboratory,
or can be  found  in nature.
Particle accelerators  can be seen as instruments for the construction
of  extreme systems  (composed of few very high energy particles)
to  study the properties of interactions at very small distances.
The history of astrophysics  can also be seen as the discovery and the study
of more and more ``exotic'' objects  (or events):
normal stars,  white dwarfs, neutron stars,
 supernova  explosions, Active Galactic Nuclei, Gamma Ray  Bursts, $\ldots$,
 whose understanding requires
a deeper and  more refined  description of the physical laws.
Cosmology \cite{Ellis} allows the study of the  early universe, 
 and   going back in time
explores  progressively more and more  extreme   conditions,  
in fact of ``arbitrary extremeness'', and constitutes the ultimate
laboratory  for fundamental science.
The three  fields of Particle Physics,  Astrophysics  and Cosmology
appear today as  more and  more  strictly interconnected  fields.

The research on Cosmic Rays  is a crucial  element in High Energy Astrophysics,
and  has a particular deep   relation  with Particle Physics.
This  relation   is historical  and methodological.
The  two  fields
started  essentially at the same time, at the beginning  of the last century,
and the measurement of the fluxes  of cosmic  rays  clearly
required, and at the same time made  possible
 an understanding of the interaction properties
of  high energy elementary particles.
Research in cosmic rays   in  the years  between  1930 and 1960 
resulted in the discovery of the 
first elementary particles (after the electron):
the  positron $e^+$, the second charged  lepton  $\mu^\pm$, 
the charged and neutral  pions $\pi^\pm$, $\pi^\circ$,
the strange particles  $\Lambda$, $K^\pm$,  
$K_L$, $K_S$   (the $\theta$--$\tau$  puzzle). 
Also the discovery of the anti--proton in 1955
happened  essentially  simultaneously 
at the  Berkeley Bevatron,  and in  emulsions  exposed to 
cosmic rays. 
Then, in the 1950's, 
 particle physics  entered in the era of  big machines  and
big detectors. An   extraordinary 
experimental  and intellectual effort culminated in the 
construction of the ``Standard Model'',   based on the
gauge group $SU(3) \otimes SU(2) \otimes U(1)$,  
with  great predictive power and a set  of 
open questions  that  have inspired  new
and very complex experimental projects such as LHC.
During the ``accelerator era''
the direction   of     scientific 
input  flowed  mostly in the   other direction, from 
particle physics to   cosmic  ray research. 
Experiments  at the ISR, the Sp$\overline{\rm p}$S and  the Tevatron colliders
measured the  interaction properties of  high energy protons,
allowing a more accurate  interpretation of the
showers produced    high  energy  cosmic rays.

Progress in  cosmic ray research  has been slow  and 
after nearly a  century of intense 
efforts, it is fair to say that  the most important questions  in 
the field  remain without  (unambiguous) answers.
The  near future  of cosmic  ray  reserach appear however
as  extraordinarily interesting, and it seems  likely that in the next decade 
we will  see dramatic advances in our understanding of the high energy
radiation.
A main reason for this  expectation is that 
cosmic  ray  research   has matured into
one   component of high  energy astrophysics,
together  with $\gamma$--astronomy \cite{gamma}, that in the last decade
has  produced  a set of  remarkable  results, and
$\nu$ --astronomy \cite{nu-astronomy} that,  after  the observations
of  the sun and SN1987a,  is now aiming at the  
detection of high energy sources using new large  volume 
$\nu$--telescopes.
Hopefully  gravitational  waves will also  be soon  observed  and we will 
receive four different  ``messengers''  ($\gamma$, $\nu$, c.r. and g.w.) 
 from astrophysical  objects.
With the combined  efforts of these   four fields, the
identification  and  detailed understanding of sources of 
galactic  and extra--galactic  cosmic rays appears possible
after nearly a century of efforts.

Cosmic ray measurements are also giving  (controversial) indications
 of  the existence of  unexpected phenomena.
The most interesting   results  is the   suggestion that there
are significant fluxes  of particles
with energy as large as several times  $10^{20}$~eV, in contrast with the
expectation that particle of such high energy cannot propagate
 for long distances.
If this  result  is   confirmed by  the future  experiments
(and we will soon know) the consequences can be   extraordinarily deep.

The  series of the  ISVHECRI (Internationall  Symposia on 
Very--High Energy Cosmic Ray interaction) discuss the science
 at the  intersection of 
the two fields of  Particle Physics  and Cosmic Ray Astrophysics.
Research in cosmic  rays  is  offering  to particle physics some 
exciting (but  again controversial) hints 
of  ``new physics'' 
such as the  existence of centauros  events
  and  of particles above the GZK  cutoff;
on  the other hand it also has  some   ``requests'',
 addressed in particular  to the
community   of physicists working on the hadron  colliders.  The  request 
is to measure  at accelerators  (and especially at the LHC)
the   main features of the very high energy
hadronic collisions, in order to interpret accurately the present and
future data  on  the  highest energy hadronic  showers.
This is  in fact a difficult and costly experimental challenge,
but  the  motivations are  strong and clear.

\section{Cosmic Ray Measurements}
\begin{figure}[htb]
\begin{center}
\epsfig{file=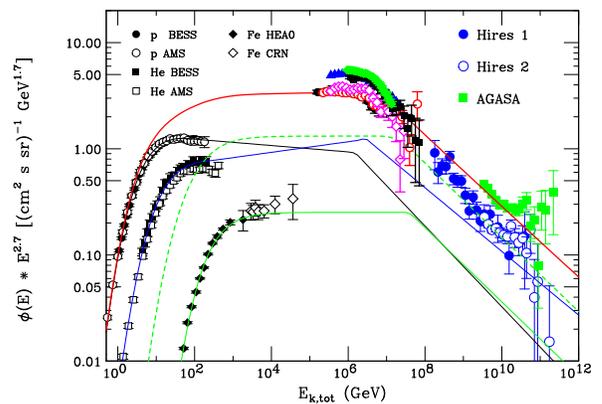,angle=90,width=7.5cm}
\caption{\small Some recent measurements of the c.r. spectra.
 The $p$ and He spectra
were  taken in june 1998. The lines are extrapolations 
of fits to the direct  measurements  \protect\cite{fit-cr}   using the
ansatz (\ref{eq:knee}) for the knee.}
\label{fig:all}
\end{center}
\end{figure}

In fig.~\ref{fig:all}  we show some measurements of the  energy spectrum
of  cosmic rays.  One can  identify  several  energy regions: \\

\noindent
{\bf [i]}
 A lower energy region ($E \aprle 30$~GeV)  where the energy spectrum
is not a simple
power law but has ``curvature'' in a log--log plot.
In this   region the fluxes of c.r.   have a time  dependence
due to  modulations  produced  by the time varying solar wind intensity.
The new measurements with  magnetic  spectrometers  
have  reduced  significantly the uncertainties
of the flux below 100~GeV,   and measurements  taken at different times
allow to    study the solar modulation, extracting  the  interstellar flux.
\begin{figure}[htb]
\begin{center}
\epsfig{file=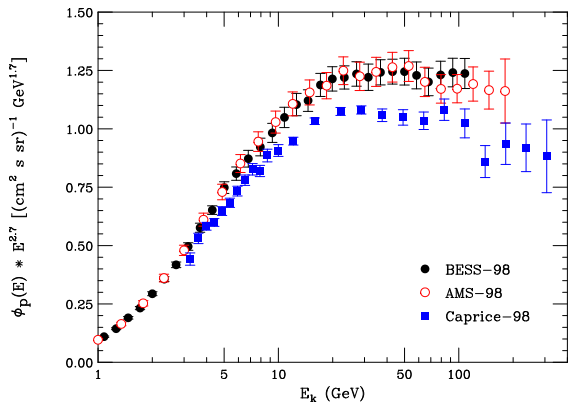,angle=90,width=7.5cm}
\caption{\small Recent simultaneous measurements of the $p$ spectrum
by magnetic  spectrometers.}
\label{fig:prot}
\end{center}
\end{figure}
There is still a  significant   difference of  order 15--20\% between 
quasi--simultaneous measurements  performed 
by the BESS \cite{bess} and AMS \cite{ams}
detectors (with higher flux)  and CAPRICE \cite{caprice} (lower flux) see fig.~\ref{fig:prot}  that  need to be  resolved.

\begin{figure}[htb]
\begin{center}
\epsfig{file=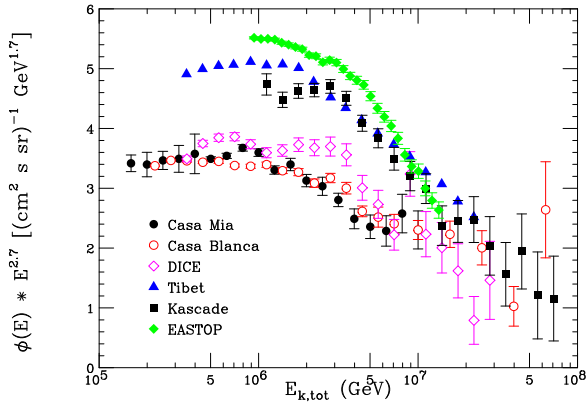,angle=90,width=7.5cm}
\caption{\small Recent measurements of the c.r. spectrum at the knee.}
\label{fig:knee}
\end{center}
\end{figure}

The main goal of the AMS detector \cite{Pohl}  is 
the  search for anti--nuclei  in the cosmic ray fluxes. The discovery of these
particle would clearly be of profound significance for
both astrophysics  and particle physics \cite{Bugaev}; the detector
 will soon start three years of data--taking aboard the International
Space Station, using a high field  superconducting magnet, obtaining data
of  unprecedented  accuracy. 
The  detailed  study of the shape of the  energy fluxes
of different  particle species ($p$, nuclei, $e^\mp$, $\overline{p}$)
in the this low energy
region has the potential to give very valuable information about the
injection, acceleration and  galactic  and solar environment propagation
of  the cosmic rays.
\\[0.2cm]
{\bf [ii]}
 In the region  ($3 \times 10^{11}~{\rm eV} \aprle E \aprle  10^{15}$~eV)
the cosmic  rays  fluxes  to a good approximation are described 
by a  simple  power law ($\phi_A (E) \propto E^{-\alpha}$). 
In this region there are only few measurements
mostly  obtained with calorimeter  on balloons,
such as JACEE  \cite{jacee} and RUNJOB \cite{runjob}. 
 There are some indications that
index $\alpha$ of the spectra of different components 
differ  and  in particular that the helium  spectrum
is  slighly harder than $p$ one  ($\alpha(p) > \alpha({\rm He})$).
 This is an important point and  need to  be confirmed  by new  
more precise measurements. 
Data of an upgraded  version of the BESS detector (BESS--TeV) should 
soon become available  possibly resolving this question.
\\[0.2cm]
{\bf [iii]}
  At the  so called   ``knee''  (at $E \sim 3 \times 10^{15}$~eV)
the  all--particle spectrum  steepens, with a change in slope
$\Delta \alpha \simeq 0.35$.
The measurements of the  spectrum   in this  region are
only obtained with indirect measurements\footnote{Clearly
an important  direction of  progress is to push  
the direct measurements  to the highest  possible energy, approaching the knee.
Ultra long duration  (60--100 days) ballon flights  in the Antartics 
offer this possibility.
The Cream detector  \cite{cream} is designed for this purpose.
}.
 A  subset of recent measurements is shown
in  fig.~\ref{fig:knee}, where we can see that significant discrepancies
exist among the  different measurements. It is still a matter of debate 
how much of the differences is due to  experimental systematic
errors, and how much is due to uncertainties  in the modeling
of the  shower development.
In fig.~\ref{fig:all} the different lines are the extrapolation of a fit
 \cite{fit-cr}   of    direct measurements  of the cosmic  ray fluxes.
There is some tension  between the results  of  such extrapolations
with the highest estimates of the flux  in the knee region by 
EAS experiments.

A significant amount of energy has  gone into the determination of the 
mass composition of cosmic rays below  and above the knee.
Perhaps the simplest model  for such an evolution is 
the assumption  that the knee  corresponds to a fixed  value of the 
rigidity  $p/Ze$, and therefore  for the  nuclear component of
electric charge $Z$:
\begin{equation}
E_{\rm knee} (Z) = Z \, E_{\rm knee} (p).
\label{eq:knee}
\end{equation}
Equation (\ref{eq:knee}) is predicted in a very  wide range  of models, where
the knee is  the  consequence of  the rigidity dependence
of the acceleration rate in the sources,
 or  the galactic  containement properties of cosmic rays.
The ansatz (\ref{eq:knee})  is  used in the    extrapolation of  the  spectra 
shown in fig.~\ref{fig:all}. 
There  is mounting evidence \cite{kascade,castellina}
that the average mass of cosmic rays  increases with energy  across the knee,
and more precisely that  equation (\ref{eq:knee}) is  valid, however large
systematic uncertainties are still  existing, and are  mostly  due
to  uncertainties in the modeling of cosmic ray interactions.

\begin{figure}[htb]
\begin{center}
\epsfig{file=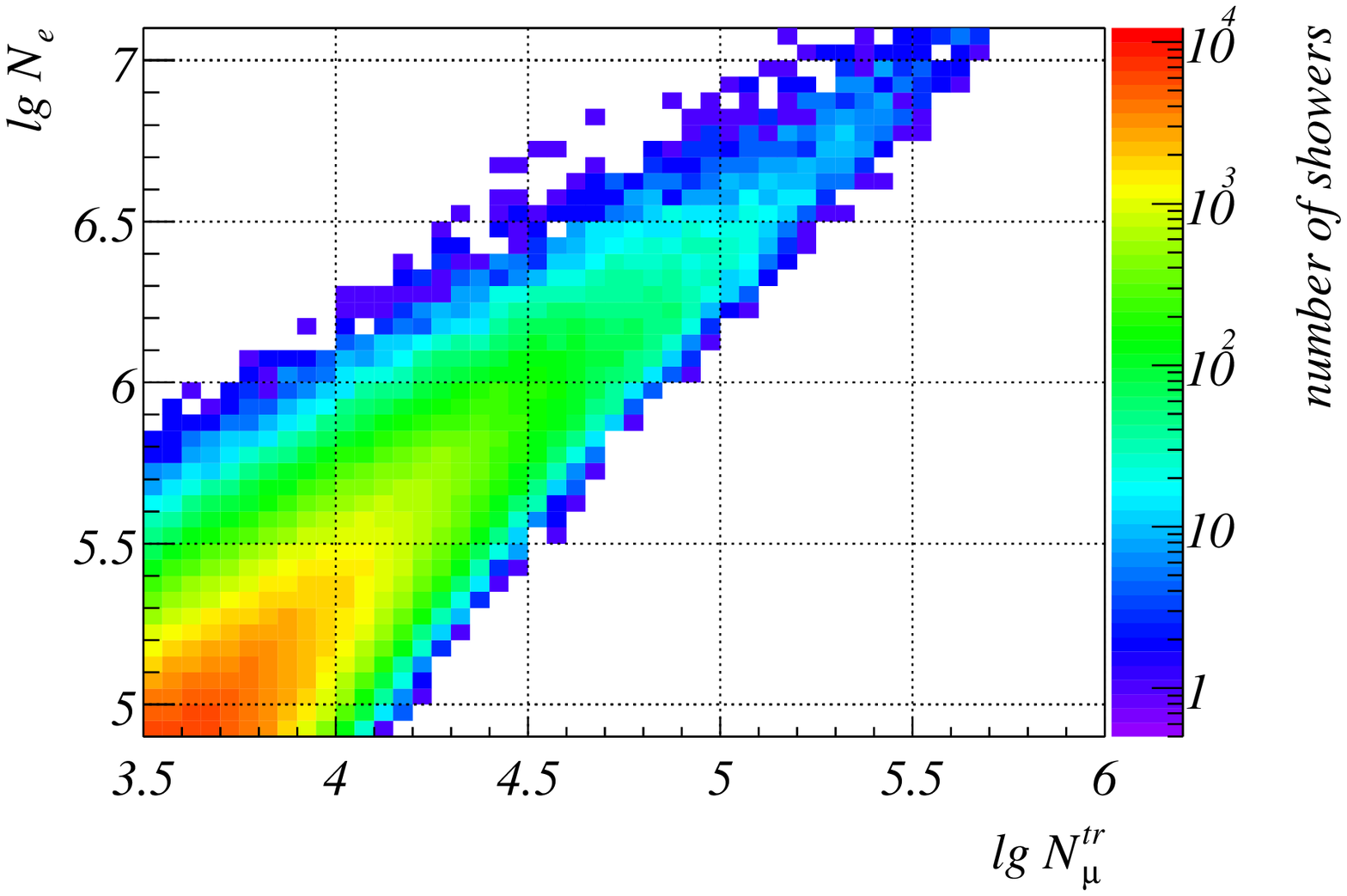,width=7cm}
\vspace{0.85 cm}
\epsfig{file=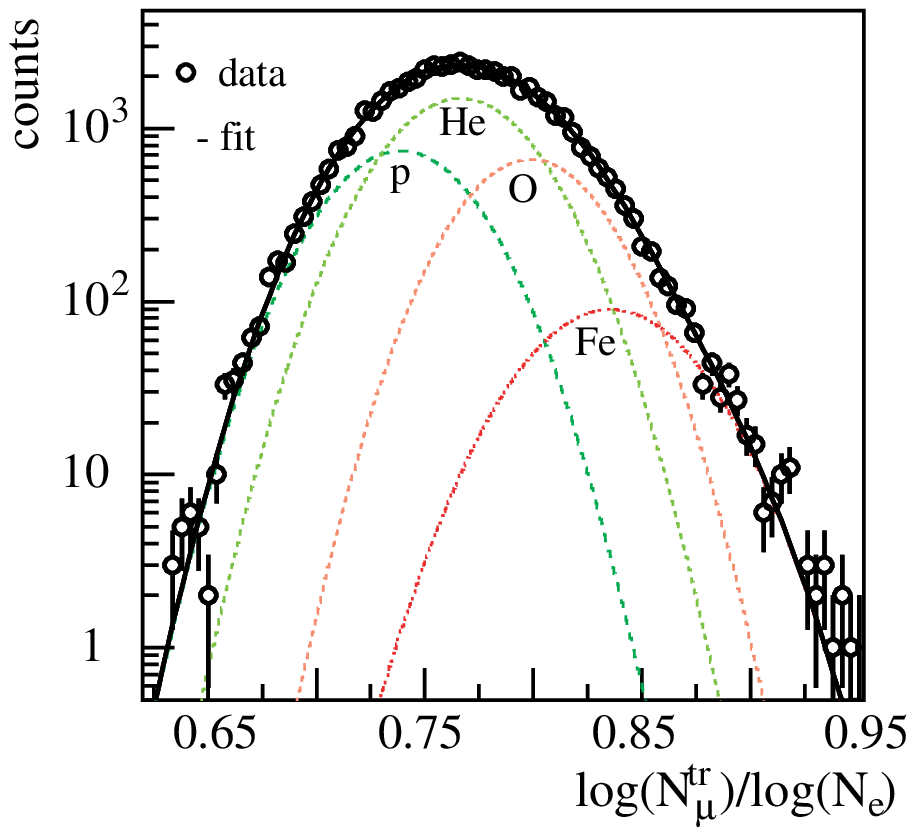,width=6.5cm,height=3.5cm}
\caption{\small The top panel  shows the 2-D  distribution
in $N_e$, $N_\mu$ of Kascade \protect\cite{kascade}.
The bottom panel  show an  example of a  composition fit.}
\label{fig:comp-knee}
\end{center}
\end{figure}

In very simplified terms the   determination of a  the mass composition
is obtained  with the measurement of 
(at least) two  quantities per shower, such as the 
electromagnetic size  $N_e$  and the  muon number $N_\mu$.
These quantities  have different  dependences
on the    energy and  mass  of the primary,  and  therefore
this allows in principle to obtain estimates of  $E$ and $A$ 
for each shower.
For example,  for the case of $N_e$ and $N_\mu$
 qualitatively  one has:
\begin{equation}
\begin {array}{lcl}
N_e \simeq K_e \; A \; \left ( {E \over A} \right )^{\alpha}
&  ~~~~~   &  {\rm with} ~\alpha  > 1 \\
N_\mu \simeq K_\mu \; A \; \left ( {E \over A} \right )^{\beta}
&  ~~~~~   &  {\rm with} ~\beta  < 1. 
\end{array}
\label{eq:ne-nmu}
\end{equation}
The esponent $\alpha$ is larger than unity because
with increasing energy the  $N_e$ size at maximum
grows  linearly with energy  while the shower maximum  
position approaches  the detector 
level, while  $\beta$ is less than unity  because 
muons are produced in the decay of mesons
 in processes such as $\pi^+ \to \mu+\nu_\mu$, 
and the decay  probability of  high energy mesons is reduced because
of the Lorentz  time expansion.
One can use equations (\ref{eq:ne-nmu}) to  express the muon number  as  a 
function of the $N_e$  and  the unknown  mass 
$A$  as:
\begin{equation}
N_\mu \simeq K^\prime \; A^{1 - \beta/\alpha} \, N_e^{\beta/\alpha}
\end{equation}
with a mass dependence $A^{1-\beta/\alpha} \sim A^{0.2}$, 
and the heavy primaries can in principle  can  be  selected choosing
muon rich showers. 
An example of this  is shown in fig.~\ref{fig:comp-knee} from 
the Kascade air shower experiment \cite{kascade}.  The detector
can measure simultaneously  $N_e$ and $N_\mu$.
In   the bottom panel of fig.~\ref{fig:comp-knee} the
 showers are selected in a fixed  interval of
$N_e$, and  the distribution in $N_\mu$  is analysed to obtain
the mass composition. 
Showers with a small   muon number $N_\mu$ are  associated 
 to proton primaries,  while the highest  $\mu$ multiplicities are associated
with iron nuclei.  A quantitative  analysis  clearly requires a precise
knowledge (including fluctuations)  of the  shower  properties 
  for primaries of different energy and mass.
The results  of fig.~\ref{fig:comp-knee}   have been fitted, using the
QGSJET model \cite{Heck},
 with  a composition   dominated by helium  nuclei and  smaller
contributions of $p$, $^{16}$O and ${}^{56}$Fe.
 It can be seen that
 the  resolution in the  measurement of  A is not sufficient to 
separate the different components,
 and therefore
the determination  of the mass composition 
depends  critically  of the  Montecarlo prediction,  and
one needs  to consider  a systematic error in the estimate  of
 the energy spectrum and mass composition  due to theoretical  uncertainties
 in the modeling of shower development (that is in the description of 
the hadronic  interaction properties).
Similar considerations  apply also  to all other  techniques for the 
determination  of spectrum and  composition in the knee region and above.
For example the DICE experiment \cite{dice} measures 
with two  imaging  telescopes  the Cherenkov light
produced by c.r. showers,  obtaining  two quantities  per shower, 
then total number of Cherenkov photons $N_{\rm Cher}$ and   the position 
of Shower maximum $X_{\rm max}$, showers  with  deep (shallow)
 $X_{\rm max}$ are
attributed to protons (iron nuclei). The BLANCA detector  \cite{blanca}
 operating in
1997--1998  measured the  distribution of Cherenkov photons at the ground 
with a system of 144  angle integrating photon detectors, extracting two
parameters  per shower, the photon density at 120 meters 
 from the shower axis ($C_{120}$) 
and the exponential slope $s$ of the photon density
in the 30--120 meters range  ($\rho(r) \simeq K \, e^{-s r}$); steep
(flat) slopes
correspond to light (heavy primaries). The energy spectrum and composition 
 can be obtained from the analysis of the distribution of events in the
($N_{\rm Cher}$,$X_{\rm max}$) or
($C_{120}$,$s$) planes\footnote{The ``unfolding''  of the spectrum and
composition from the data  is a non--trivial   statistical problem,
even under the assumption of no--systematical  errors, and several
approches are possible. See for example
the contributions of Kascade at this  conference \protect\cite{kascade,Roth}.}.
\begin{figure}[htb]
\begin{center}
\epsfig{file=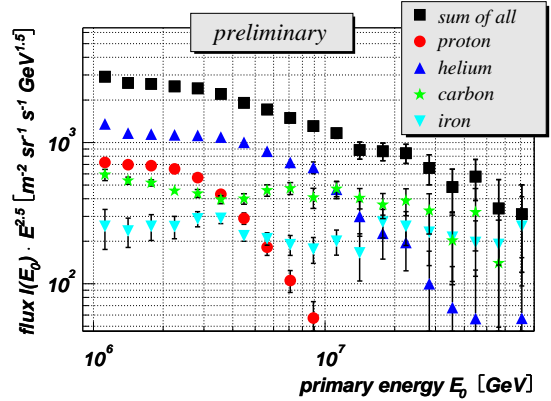,width=7.5cm}
\caption{\small Fit of the composition at the knee  obtained
by  Kascade \protect\cite{kascade}.}
\label{fig:comp-kascade}
\end{center}
\end{figure}

Some  detectors  can measure more than two quantities per shower, for example
Kascade  can measure    not only the  electron and muon sizes
($N_e$ and $N_\mu$), but also  the   hadronic  component $N_{\rm had}$ in
its central calorimeter \cite{Mielke}. For the a  fixed  energy,
  light primaries showers are more penetrating,
 and the hadronic  component is larger.
The analysis of the data in terms  of  different  pairs of variables,
for example ($N_e$, $N_\mu$) and 
($N_e$, $N_{\rm had}$),
will give consistent results   only  if the modeling of the shower development 
is correct.  Similarly 
 (barring  the existence of experimental systematic errors)  the interpretation
in terms of spectrum and composition  of different experiments will  
be compatible only if the  modeling of hadronic  interactions  is sufficiently
accurate.  This requirement of {\em consistency} (within and 
 between experiments) allows   in principle to obtain  at the same time 
information about the spectrum  and composition of primary c.r. {\em and}
about the properties of hadronic  interactions.
This {\em bootstrap} philosophy has been at the center of considerable  efforts
in recent years  (se for example  \cite{Erlykin}   for a contribution
at this conference). A critical analysis of all available data is  beyond the 
scope of this  summary
(see \cite{Swordy,Kampert}  for a review and critical analysis). In a nutshell
the main points are the following:
(a) significant inconsistencies still exist  within and between experiments,
pointing to the necessity  of an improved modeling  of hadronic  interactions;
(b) a consistent picture is however beginning to emerge, the   existence of
the ``knee'' is firmly established, even if the precise shape and location
are still uncertain may be by a factor as large as two, and most   experiments
extract a composition that becomes  heavier  across the knee, in agreeement
with  the assumption of the knee as a  rigidity dependent 
 feature (see fig.~\ref{fig:comp-kascade} 
for an example)\footnote{There are however significant discrepancies between 
experiments.  Analysis of the Cherenkov data  suggests a
composition becoming  lighter. See \protect\cite{Swordy} for
a critical discussion.};
 (c) the general features  of the  hadronic interactions incorporated in the
Regge--Gribov models currently in use are at least  qualitatively correct.
 
\begin{figure}[htb]
\begin{center}
\epsfig{file=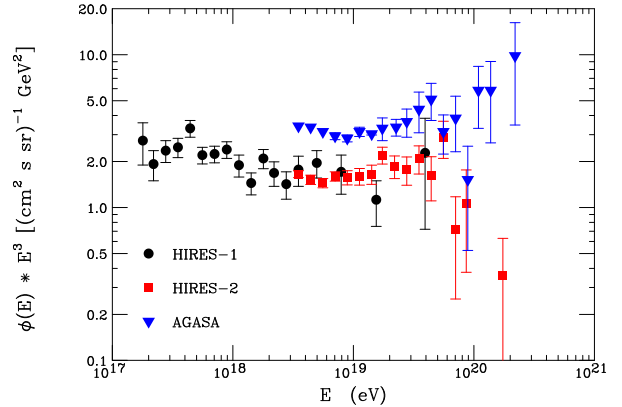,angle=90,width=7.5cm}
\caption{\small Recent measurements of the c.r. spectrum at the
highest energy.}
\label{fig:uhe}
\end{center}
\end{figure}

The energy region above the knee  ($10^{16} \le E \le 10^{18}$~eV), is still
relatively poorly  known. The Kascade--Grande  detector is planning
to explore it, with the main aim to  identify an ``iron knee''  at an
energy $E \sim  6 \times 10^{16}$~eV \cite{Kampert-Grande}.
A detailed analysis of the size spectrum of 
7 different  air shower arrays  \cite{Schatz}
already 
gives some  qualitative indications  of the existence of a second knee,
that could be  attributed to  the bending of the iron component.
Some authors \cite{wolfendale}  see evidence  as a more detailed  structure
in the knee energy spectrum,  that are attributed to the contributions
of a recent nearby supernova explosions.  \\[0.2cm]
{\bf [iv]}
 The  highest  energy points  in fig.~\ref{fig:all} and~\ref{fig:uhe} 
 are from  the Agasa \cite{agasa}  and Hires
\cite{hires} detectors, the data  of the Yakutsk array can be seen 
in \cite{yakutsk}.
  The Agasa spectrum extends up to an  energy 
$E \sim 3 \times 10^{20}$~eV.
It is well known \cite{nagano}  that   one expects the
existence  of a  (Greisen--Zatsepin--Kuzmin or GZK) cutoff
in the energy spectrum of cosmic rays  due to 
 interactions  with the cosmic  microwave background.
The dominant process  is  pion  photoproduction  on the
photons of the  ($2.7^\circ$K) Cosmic Microwave Background Radiation:
$p + \gamma_{\rm CMBR} \to p(n) + \pi+ \ldots$, with 
an energy threshold  of order $E_{\rm thr} \simeq m_p \, m_\pi /\langle 
\epsilon \rangle \simeq {\rm  few } \times  10^{19}~{\rm eV}$. 
Particles  above  the GZK  cutoff should only arrive  from
near  (on a cosmological case) sources.
Since the $\gamma$  target is  very precisely 
known  
%(the contribution of  non--cosmological photons
%generated after  galaxy formation is small and only  increases the  absorption)
and the interaction  cross section has been 
accurately measured  in experiments   with protons at rest, it is 
possible to  compute  with very good precision the  interaction
length and  energy loss  of Ultra High Energy (UHE) protons. 
Similar considerations can also be made for  composite nuclei,
when the   dominant energy loss process  is photodisintegration
(such as $A + \gamma \to (A-1) + N$).
The detailed shape of  UHE cosmic rays  flux  will depend  on the 
shape of the spectrum at the source (in particular on the 
maximum acceleration energy $E_{\rm max}$), 
the distribution  in space--time of the sources, and the  
structure of the  extra--galactic magnetic fields, that control the
propagation  of   charged 
 particles   from  the source  to our galaxy.

The energy determination of an EAS  detector as Agasa (see \cite{Takeda}
for a full discussion)
 is based on a measurement of the  particle  density at the ground  
at a  distance $\sim 600$ meters from the shower core.
It has    been   demonstrated \cite{Hillas}  that this measurement  
is   relatively insensitive to
both the mass of the primary particle and the details of the interaction
model,  however in principle some model  dependence is possible and needs
to be very carefully investigated.

New results of the High Resolution Fly's Eye, based on the fluorescence
technique \cite{hires} 
 show a spectrum that is well described assuming the existence  of the
GZK cutoff.
In principle the measurement  based of the detection of fluorescence
ligh emitted by  nitrogen molecules  excited by the shower,  represents
a nearly completely model independent   method for  the energy determination.
If $N(X)$ is the number of charged  particle in a shower  at  depth $X$  and
$dY_{\rm fluo}/dE$ is the yield   of fluorescence  photons produced   
after  the release of  the energy $dE$ in ionization, then   the  number of
fluorescence photons  generated by the shower   in the depth
interval  ($X$, $X + dX$) is:
\begin{equation}
{dN_{\rm fluo}  \over dX } = N_e (X) \, 
\left \langle -{dE\over dX}  \right \rangle 
\; {dY_{\rm fluo} \over dE} (X)
\end{equation}
the $X$ dependence of the yield   reflects a (strong) 
dependence on the air pressure.
The number received   at the detector, can be  obtained 
from  simple  geometry 
(the fluorescence  emission is isotropic) 
and  a knowledge of  the shower  axis position 
if one has good  control of the
trasparency of the atmosphere for the  fluorescence photons.
This technique in principle allows to  measure  the profile  of the
fluorescence  emission along the shower axis, and therefore the
profile of the shower energy loss, and by integration
the total  energy   dissipated in ionization  by a c.r. shower.
The   total energy of the primary particle can then
be  obtained  applying  small corrections
for the  energy that reaches the  ground in the form of  neutrinos,
muons, hadrons  and the tail of the electromagnetic  shower.
In this case the main sources of systematic  uncertainties  are the 
correct description of the fluorescence yield and of the 
atmospheric  transparency.

\subsection{Particles beyond the GZK cutoff ?}
The question of the existence of a significant  flux
of particles particles with energy above the expected GZK cutoff
is certainly the question thas has attracted more  excitement and controversy
in cosmic ray physics in the last decade.
In the following there is a list of possible solutions for this
puzzle: \\
(i)  There are no  particles  above the  GZK cutoff. The present 
results of Agasa, Fly's Eye, Haverah Park and othe detectors 
are  the  effect of a combination of incorrect energy calibration,
larger  than predicted fluctuations in  shower development, non  gaussian 
tails in mesasurements etc. \\
(ii)  The highest energy particles are
produced in few ``standard''  sources   at small distances.
The energy  spectrum  and the 
angular  distribution expected in this  scenario
will then have strong features that should be 
clearly demonstrated
with the higher   statistics  obtained  from future experiments. \\
(iii) The highest energy particles are generated  in the vicinity of
our galaxy  by the interactions of a  more weakly interacting
``carrier''.  This is  the so called  ``$Z$--burst''  scenario \cite{Fargion}
  where the carriers  are  ultra--high energy neutrinos. These neutrinos
are created in ``standard'' sources, propagate in intergalactic  space
with negligible  absorption, 
produce the observed  UHE particles   interacting  with  a
postulated  neutrino  galactic halo.
This scenario    implies    that  there are neutrinos  with 
energy $E_\nu$ at least a few  times
$E_{GZK}$   (and therefore at the source
 protons must be accelerated up to an energy
several  times   $E_\nu$).  To obtain a sufficiently large  
probability for interactions with the galactic halo
this scenario  assumes   that the cross section  is  enhanced by 
the  $s$--channel resonant   production of $Z^0$ 
(the process $\nu +  \overline{\nu} \to Z^0 \to {\rm final~state}$).
This  requires a  neutrino mass of order
\begin{equation}
m_\nu = {M_Z^2 \over 2 E_\nu }  \sim 
{M_Z^2 \over ({\rm few}~10^{21}~eV) } \simeq 
{\rm  few ~eV}
\end{equation}
that could be the correct number  consistent with the existence
of the halo. \\
(iv) The existence of  UHE  particles,    coming from 
``invisible''  sources  can be  naturally explained  in the  framework
of the so called Top--Down \cite{Top-Down}  models. 
In these models  the   UHE particles are not  produced  by acceleration
(the   Down--Top  mechanism)   but are the result of the {\em decay}
of  very large mass particles. The  mass scale
$M_X$  is related   to a unification mass scale.
In particular it is  important  to note
that the  grand--unification mass  scale
($M_{\rm GUT} \sim 10^{24}$~eV)  is of the right order  of magnitude
so that  the decay   of particles with mass $M_X \sim M_{\rm GUT}$  
can produce the super--GZK  particles.
The Top--Down models   require that the dark matter in the universe
is  provided  by  such super--heavy particles, or by   topological  defects
(such as  magnetic monopoles, cosmic strings, $\ldots$)    that can 
decay into such particles. \\
(v) Finally one extraordinary possibility is the existence of
violations of Lorentz  invariance. 
In this case  the  GZK cutoff   is  not observed 
because  it does not exist !
 The statement   that 
protons  of energy $10^{20}$~eV  
produce pions interacting with photons  of $\varepsilon \simeq 10^{-3}$~eV
is based on the observations  of the  interactions properties 
of   photons  with $\varepsilon_{\rm r.f.} \sim 10^{8}$~eV   with protons
at rest, and on the assumption of Lorentz invariance  and the  validity
of Lorentz invariance. The  two frames are connected by a transformation
with a Lorentz $\gamma$  factor of order 10$^{11}$.
If Lorentz  invariance  is violated the statement can become  false.
This  apparently outrageous possibility  is  actually  {\em predicted}
\cite{Amelino-Camelia} 
in the framework of quantum gravity  or in  models  where  the space
manifold  has additional  large extra  dimensions.

It should be possible to  determine which one   is the true
true solution of this puzzle  thanks to the  new detectors in
construction (as Auger \cite{Dova})  or in the planning stage
(as Euso \cite{Scarsi}, based  on the detection of 
fluorescence light from  space.)

\section{Cosmic Ray Astrophysics}
A list of  fundamental questions for c.r.  astrophysics
can be simply formulated as following: \\
(A) What if the dominant source  for cosmic  rays 
below the ``knee''  ? \\
(B) What is  the origin of the knee ? \\
(C) What is the origin of the particles  beyond the knee ? \\
(D) At what  energy the    fluxes of
  extra--galactic and galactic  cosmic rays  are equal ? \\
(E) Which are the sources of extra--galactic  cosmic rays ? \\ 
(F) Are there particles  beyond the GZK cutoff ? \\

It is surprising that  we still do not have unambiguous 
 answers to any of these questions.
There is a general (but non universal)  consensus  that 
SuperNova Remnants  (SNR's)  are the source  of the galactic  cosmic
rays. This consideration is  essentially   based on  two considerations:
(i) SNR  can  provide the power ($L_{\rm c.r.} \sim 10^{40}$~erg/s)
 needed to mantain    the observed 
energy density of cosmic  rays,  taking into account  the  measured 
(rigidity dependent) confinement time  of the cosmic rays;
(ii)  the  mechanism of diffusive  first order   Fermi acceleration
at a shock   can  naturally produce a power law source energy spectrum
($q (E) \propto  E^{-\alpha}$)
with a (differential) esponent $\alpha \simeq 2 + \varepsilon$ 
(with $\varepsilon$ small).
 
In  diffusive  shock acceleration a   charged  particle   moving in 
a turbolent magnetic  field  of average strength $B$
performs  ``cycles''  crossing back and forth  across the
shock discontinuity.  During each cycle the particle  acquires an energy
$\Delta E \simeq  {4/3} ~\beta_{\rm sh} \, E$
(where $c\, \beta_{\rm sh} = v_2 - v_1$ is the difference
between the velocity  of the fluid on the two sides of the shock.
The time for performing a cycle  is  of order
$T_{\rm cycle} \simeq D/ (\beta_{\rm sh} \, c^2)$, where $D$ is the
 diffusion coefficient  that depends on the intensity and structure of the
magnetic  field.
For a  diffusion coefficient   linear in $E$:
\begin{equation}
D \simeq {1 \over 3} \, r_L  \, c \simeq {1 \over 3} \, {E \over Z e B} \, c 
\end{equation}
the acceleration rate  $dE/dt \simeq \Delta E/T_{\rm cycle}$
   becomes  a constant, and the
maximum energy  obtainable  is a Supernova   can be estimated  
(using $R_{\rm SNR} \sim c \, t_{SNR} \, \beta_{\rm sh}$) as:
\begin{equation}
E_{\rm max}  \sim  {dE \over dt} \; t_{\rm SNR} \sim
 R_{\rm SNR} \times Z \times  B \times \beta_{\rm sh}
\label{eq:emax}
\end{equation}
This  energy is   similar to the energy of the knee. 

The simple ``standard'' scenario  outlined  above predicts
an exponential cutoff   of the cosmic ray flux  at the maximum energy.
This is very different from the simple  moderate steepening of the 
energy spectrum observed  in the data,
 and therefore the 
identification of the knee with the maximum energy in SNR remains
unclear.

A problem  that still waits for a clear answer is the determination of the
energy at which the fluxes of galactic  and extragalactic particles
are equal.   There is a general  consensus that this  energy must exist.
The gyroradius of   charged particles in a magnetic  field is
\begin{equation}
R_{\rm gyro} = {p_\perp \over Z \, e \, B} \simeq 1.1 ~{ p_{\perp} (10^{18}~
{\rm eV}) \over Z \, B( \mu {\rm Gauss} ) }  ~{\rm Kpc}
\end{equation}
The  galactic radius  is $r_{\rm gal} \sim 15$~Kpc, and the typical strength of
the magnetic  field is $B \sim 3~\mu$Gauss,
therefore very likely 
the  highest energy cosmic ray   cannot remain  confined
inside the Galaxy.
It is possible that the  energy where the fluxes of particles of galactic and
extra--galactic origin   are equal  corresponds to the
so called ``ankle'' ($E \sim 10^{19}$~eV) 
 in the c.r. spectrum,  however it is also claimed
\cite{berezinsky} that the ankle is an asorption feature due
to the process $p + \gamma_{\rm cmbr} \to p + e^+ + e^-$ and that the
crossing point for the two population is at lower energy.
This is  clearly an important point that  can be clarified with
additional data on the energy spectrum and  angular anistropy  of cosmic rays.

Equation (\ref{eq:emax})  can be used to   constraint the size
and magnetic  field  of  any source of cosmic rays where the acceleration
mechanism is  first order  Fermi acceleration.  This  implies that very 
few objects in  the known universe  can  have  a sufficienty
large  product  $R_{\rm source} \times B$ to  be a candidate
for  the acceleration of the UHE cosmic ray \cite{Hillas1}.
AGN  and  GRB are perhaps the  best candidates  for this purpose.

\section{Hadronic Interactions}
The study of high energy cosmic rays   requires  ``indirect methods''
that is the measurement  of  the shower produced  by the primary particle.
The uncertainties  in the  prediction of the development  of  the shower
produced by a primary cosmic  ray are the consequence of  uncertainties
in the calculation of hadronic interactions.
The  problem  is the following:
one has a ``projectile''  particle (a proton, a nucleus (A,Z),  
or a  weakly  decaying   meson such as
a pion or kaon)  and one needs  to know the interaction cross section
with the air nuclei (mostly nitrogen and oxygen), and the properties of
the final  state  produced in such an interaction, namely the 
multiplicity,   flavor  composition  and  momentum distribution
of the final state particles, with a correct estimate  of  
the fluctuations.
It is well  known  that  now  and for the foreseable feature we are not
in the condition to compute from  first principles  these needed
quantities from the fundamental  QCD  Lagrangian.
Moreover the existing data 
do  not  cover all the ``phase space'' necessary for purely phenomenological
description.
The c.m.  energy  on nucleon--nucleon
interactions  for  cosmic  rays  in the knee  region
($E \sim  3 \times 10^{15}$~eV) and 
and near the  GZK cutoff energy 
($E \sim  10^{20}$~eV)   are:
\begin{equation}  
\begin{array}{lcc}
(\sqrt{s_{NN}} )_{\rm knee} & \sim &  2.5/\sqrt{A} ~{\rm TeV }  \\
~ & ~ & \\
(\sqrt{s_{NN}} )_{\rm GZK} & \sim  & 400/\sqrt{A} ~{\rm TeV } 
\end{array}
\end{equation}
($A$ is the mass of the primary particle).
The highest energy  collisions produced in an accelerator  have
$\sqrt{s_{p\overline{p}}} \simeq 1.8$~TeV at the Tevatron,  and therefore
corresponds closely to the knee; the LHC  collider  at CERN will
reach
$\sqrt{s_{pp}} \simeq 14$~TeV, that is still approximately 30 times
lower  than the GZK energy.
However the situation is   much worst than what appears  from these simple
considerations.  The measurements at the hadron colliders   have been
limited to an angular region   that  excludes the  beam pipe,  and therefore
a very  large majority of the high energy particles  that  are emitted 
at small angles    are unobservable  (see fig.~\ref{fig:eta}). 
These particles  carry more than
90\% of the energy in a collision and are clearly   those  crucial in
determining   the properties of air showers.
It should also be  noted that  the study of hadron--nucleus interactions
is still limited  to fixed  target energies
 $(\sqrt{s_{NN}})_{hA} \aprle 0.027$~TeV).
The new data from the RHIC  detector  about gold--gold detector  at
 $(\sqrt{s_{NN}})_{AA} \aprle 0.2$~TeV)
presented at this conference by S.~Klein \cite{klein}
 have  therefore  great value in
testing the accuracy  of the treatment of nuclear effects used in the existing
montecarlo codes.  In fact a comparison    has  shown the existence of
 non--trivial  discrepancies (15--20\% in the central region
rapidity density, see \cite{ranft,Engel,Werner}). 

At fixed target energies the inclusive distribution of  final  state particles
exhibit in   first approximation the property of 
Feynman scaling:
\begin{eqnarray}
{d\sigma_{pp \to a} \over dp_\parallel \, d^2 p_\perp} (p_\parallel, p_\perp, \sqrt{s})
&\simeq  & 
f_a(p_\parallel, \sqrt{s}) ~G_a (p_\parallel)  \nonumber \\
& \simeq  &
{ F_a(x_F)\over E} \; G_a(p_\perp)
\end{eqnarray}
where  $x_F =  2 p^*_\parallel/\sqrt{s}$  and 
the  functions $F_a(x_F)$ and $G_a (p_\perp)
 \propto e^{-bp_\perp^2}$ are independent
from $\sqrt{s}$.
Clearly the assumption of Feynman scaling   allows to extrapolate  the 
low energy results and to  predict  the properties of showers  of  arbitrary
energy.
However the data of the hadron  colliders (ISR, Sp$\overline{\rm p}$S,
 Tevatron and RHIC)  have shown that Feynman scaling is violated.
As an example the scaling function $F_\pi$ for pion production 
 in $pp$ interactions has  approximately  the 
 form  $F_\pi (x_F) \simeq C \,(1 - |x_F|)^n$  with $n \sim 3$--4.
   This  form   indicates that pions  are
approximately produced  with a spectrum $dn/dE \sim 1/E$   peaked  at low
energy.
At collider energy the   quantity $C$  (that is the
value  of $F_\pi$     near  $x_F \sim 0$ 
or the  height of the rapidity  plateau)  is  measured
to grow  logarithmically with   increasing $\sqrt{s}$.
The form of scaling violations for large $|x_F|$ (for nucleons and
mesons) is   known much
more poorly, however it is 
essential for  shower development.

The spectrum of the nucleons  produced in hadronic interactions
plays a fundamental role in   the development of c.r. showers.
At fixed  target energy   a fraction $\sim 20$\% 
of the $pp$ inelastic interactions
is due to ``single diffraction''  where one if the incident protons is excited 
into a state $X$  with the  same internal quantum numbers  that scatters 
elastically with  small transfer momentum with  the other proton
and  ($\sim 5$\% of the inelastic interactions can be attributed 
to  double diffraction).
Note that in a  target diffraction event   the projectile proton
retains nearly all the initial  energy, while in projectile  diffraction,
the  decay of the excited  state $X$  result in a final state nucleon
that carries a very large fraction ($\aprge 50$\%) of the  initial $p$ energy.
In non--diffractive interactions  the final state nucleons have a 
 hard--spectrum ($dn/dE \sim {\rm const}$)  and   carry  approximately
$\sim 40$\% of the initial state energy\footnote{Traditionally the energy
fraction  carried by nucleons   has beed called by comsmic
ray  physicists the ``elasticity''  of the interaction.}.
These high energy nucleons  in the final state feed energy deeper into the 
shower  and clearly play a very important role in 
the shower development.
At the hadron colliders  most  of these nucleons  are unobserved,  and 
their spectrum must be infered  with a large amount of  uncertainty.

\subsection{Montecarlo Modeling}
A general  framework  to compute the properties of hadronic 
($hp$, $hA$ and $AA$)  interactions   has  been developed in
  the last 10 years \cite{regge-gribov}.
In this  frameork, the so called ``Regge--Gribov effective theory'', that is 
formally very similar to an eikonalized parton model,  an hadronic 
collision is analysed as a set  of sub--interactions, or ``Pomeron exchanges'',
between the participant particles.
A fraction of these sub--interactions can  be simply understood as hard 
or semi--hard interactions between  partons, that can be treated 
in perturbative QCD, while another fraction is ``soft''.  The  growth of
the cross section with energy  is related to the  increase of
the number of sub--interactions  with  increasing $\sqrt{s}$.  This approach
can be naturally implemented into  montecarlo algorithms. The ``topological
structure'' of one  event, that is the number and type of sub--interaction
is  translated  into the formation of a set of  color strings (closed loops
or objects with $q$, $\overline {q}$  or $qq$ ``endings'')   conserving exactly
4--momentum and all quantum  numbers.  These strings are then  fragmented
into  observable hadrons using algorithms  similar
(or identical) to the algorithms  developed by the LUND group.
Several  montecarlo  implementation  of this philosophy
(QGSJET, Sibyll, DPMJET, VENUS, NEXUS)
 have been developed  and are 
used  in the   montecarlo simulation of c.r. showers.
A discussion of the differences  between these MC  implementations  can be
 found in \cite{Heck}.
It is  encouraging that the differences  between  the latest versions of the
 models    are   smaller than in the past.
The size of these differences has  been used to estimate the importance of
systematic uncertainties in hadronic interactions  modeling. 
 It  should be noted  that the use of  this method
  has the danger to underestimate  systematic
errors, because all of these codes   share the same  basic 
  theoretical assumptions, and
therefore  naturally converge   to similar results.
It should not be forgotten   that  despite of their   sophisticated  language 
the theoretical basis for this models  is not rock solid, and 
significant uncertainties  still exist.
In fact several  important  problems  do not have  an unambiguous answer
in the framework of the Regge--Gribov approach. In particular it is not 
clear  how diffraction fits in  the theoretical  scheme;  there is also
significant arbitrariness in the shape (and evolution with energy)
of the inclusive particle  distribution in the fragmentation regions
($|x_F| \aprge 0.1$).
It is  significant that each time  new accelerator data has  become available
(from ISR to the recent RHIC data)  significant differences
with the available predictions were  found.  New  data is clearly required
to validate (or correct)  the  existing models.

\subsection{NEEDS: Requests of cosmic ray physics to accelerator physics}
For a full and correct interpretation of the cosmic  ray shower measurements
at and above the knee   new data from accelerator  experiments are
required.  The  cosmic  ray  community has invested  significant efforts
in the ``bootstrap method''  (extracting the cosmic ray energy spectrum and
composition {\em together}  with the main features  for hadronic  interactions
purely fron c.r. shower measurements),   and important theoretical  efforts
are made to construct well  motivated extrapolations of the existing data into
the required phase space   region ($\sqrt{s}$,$x_F$, $p_\perp$); however there 
is a broad consensus in the community  that the best perspectives for
progress  in improving energy and   mass resolutions is new data from
accelerators.
A workshop (opportunely named ``NEEDS'')  was   held in Karlsruhe 
in april  2002 \cite{needs} to discuss which  measurements   of hadronic   interaction
properties are most important  for the field, and how it is possible to obtain 
them.  This   discussion continued at the ISVHECRI conference
(see \cite{Engel,Knapp} for a review and discussion).
Some central questions are:  \\
(i) How important are  the uncertainties   in our knowledge of
hadronic  interactions in the determination of the  c.r.
flux and composition ? \\
(ii) What  impact can   have the  planned  future  experiments 
in reducing these  uncertanties ? \\
(iii)  Which  additional experimental programs can help in 
further reducing  these uncertainties ? \\

A brief list of the most important measurements  for shower development
could be: \\
(i) Precise  measurements  of total and inelastic  cross section.\\
(ii) Measurements of the ratio  $\sigma_{\rm  diff}/\sigma_{\rm inel}$.\\
(iii)  Energy distribution of the  leading  nucleon in the  final  state. \\
(iv) Inclusive pion spectra    in the  fragmentation region $x_F \aprge 0.1$.\\
The optimum would  clearly be to have these measurements
for $pp$, $pA$  and $AA$ collisions at the LHC collider.

\section{Emulsion Chambers results}
\begin{figure}[tb]
\begin{center}
\epsfig{file=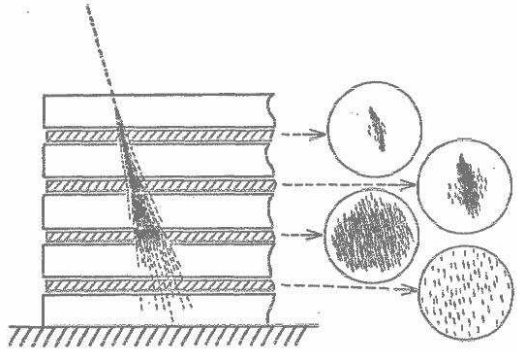,width=7.0cm}
\caption{\small Simplified scheme of an Emulsion Chamber
(from \protect\cite{Lattes}).
}
\label{fig:emuls1}
\end{center}
\end{figure}

\begin{figure}[htb]
\begin{center}
\epsfig{file=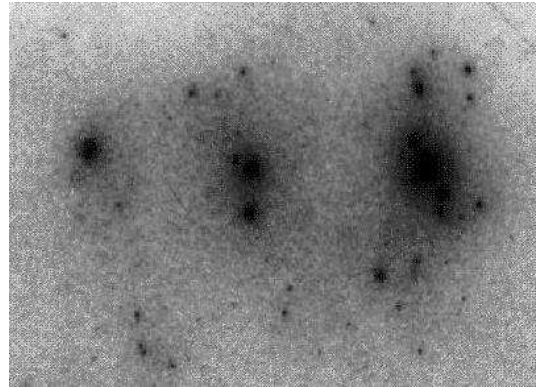,width=7cm}
\caption{\small Example of one ``family''  in a 
sensitive layer of an emulsion chamber. This  is part of the
Centauro--1 event.}
\label{fig:emuls2}
\end{center}
\end{figure}

The Emulsion Chamber technique, developed in the Chacaltaya laboratory
in Bolivia ($h=5200$~m)   has been in use for more
than 30  years in laboratories placed at mountain  altitude
\cite{Lattes,Slavatinsky}. 
The basic  structure of an emulsion  chamber  is a sandwich  of
absorber  (lead)  layers   alternated  with  sensitive (emulsion or
X--ray film) layers  as  illustrated in fig.~\ref{fig:emuls1}.
A chamber (with typically a surface of order 10~m$^2$)  is   exposed 
for a time interval of several months, then  the sensitive layers are removed,
developed and analysed.
A charged  particle crossing a sensitive layer  leaves 
a track that is visible with a microscope;
a shower  composed of many   nearly parallel and  closely packed particles
leaves a dark spot. The darkness is 
measured with  photometers analysing  the transparency of the 
sensitive layer around that position. Spots  corresponding  to the same
shower can  be associated with each other 
obtaining  a  `longitudinal darkness profile' for a shower, and from it
an estimate of its electromagnetic energy.
The detection  threshold for a shower depends on the 
sensitive material used and the level of background present, and is typically
$\sim 1$~TeV. At this energy the spot of a shower at maximum is
visible with naked eyes, and the scanning process is much simplified.
The  interaction length  in lead  ($\lambda_{\rm int} \simeq 18.5$~cm) 
is  much  longer that the radiation length ($X_0 \simeq 0.57$~cm), 
therefore an  emulsion   chamber  effectively  measures only
the electromagnetic component of a shower.
Because of the large difference  between $\lambda_{\rm int}$ and $X_0$
photons  and charged  hadrons   arriving to the chamber can
be  clearly separated, since  $\gamma$--induced showers  initiate
after  one radiation length (in practice  after the first 
absorption layer), while hadron showers   initiate  after 
one  interaction length (that is  several absorber layers). 
In  many  chambers, to enhance the hadron detection efficiency, the emulsion
chamber is  divided into two  parts    with in the middle a low $Z$ material
(for example  carbon)
 target layer   where  hadrons can interact generating photons
(via $\pi^\circ$ decay)  that are then  detected in  the lower  chamber.
 For hadron--induced showers 
only   the energy fraction that goes into  $\pi^\circ$ production in the first
interaction ($\sim 0.2$ of the initial   energy  for a $p$) 
is  visible in the chamber.
A primary particle  interacting above the detector  will   produce
several secondaries, the high energy ones  that  reach the emulsion chamber
generate  a  bundle  of close spots due 
to the quasi--parallel showers  that are collectively called
a ``family''  (see  fig.~\ref{fig:emuls2}).  The showers of a family 
can be analysed together to obtain information
about the nature of the primary particle, and  about its
interaction properties.
In some cases  it is possible  to deduce the position of the
primary interaction point  by  triangulation  as the point where
the sub--shower axis converge.
Clearly the emulsion--technique is a  very interesting  method to 
study with   very fine  resolution the  core of high energy cosmic ray showers.

For  a long time  there  have been claims of the existence of unusual
events  in  the emulsion chambers.  The most well known and most interesting
type of these ``exotic''  events are the ``Centauros'' 
\cite{Lattes,Baradzei,centauro}.
A ``Centauro'' is an  event  (a family)  
 here   a large fraction of the total  visible energy
is attributed to  hadron showers.
For example the celebrated Centauro--1  event  detected  30 years ago  in
Chacaltaya  laboratory \cite{Lattes}
 was  composed of  49 hadron showers and only one $e/\gamma$ shower. 
A small number of Centauro  candidates  has  been obtained
by the Chacaltaya  and Pamir collaboration\footnote{The number  of candidates
depend  on the exact  definition of ``centauro''. 
Experimentally centauro candidates  fall in    the region
of large $N$  and large $Q_h$. 
in the plane
($N$,$Q_h$)   where $N$ is the number  of  showers in a family
and $Q_h = E_h/(E_h + E_{e/\gamma})$ 
is the  fraction of the reconstructed energy   attributed
to hadrons). 
Depending of the  cuts applied  the world sample of Centauro 
candidates  
is of order $\sim 5$--10.
It has  not been  clearly  established 
if they  constitute a  distinct population or are the tail of
a single  distribution  that fills the entire  $(N,Q_h)$ plane.}
 while no  events have been found
at Mount Fuji and and Mount Kambala. 

Thirty  years of investigations have not  clarified the 
nature of the centauro phenomenon.
Possible interpretations  fall into  two categories:  exotic primaries
(such a compressed glob  or hadronic matter \cite{Bjorken-centauro} or
a strangelet), or the result of new  features of hadronic  interactions
at high energy, such as the formation
of ``disoriented chiral condensate''.
Also the  possibility 
that  the  events have an  explanation 
in terms of  standard physics,
fluctuations and selection effects cannot  be disregarded.
All type  of  explanations
run into  significant  difficulties.
 Searches  for  centauro--like events 
at accelerators \cite{centauro-ua5,centauro-ua1}, including the 
recent  RHIC data have been negative.
  These negative results cannot completely eliminate the
 interpretation  of  Centauro events as a  feature of the  hadronic
interactions, because  not the entire  relevant phase space 
has been  covered. 
A new  specialised  detector CASTOR   (within CMS) \cite{castor}
(covering the   very forward  pseudorapidity region $5.5 \le \eta \le 7$) 
has been proposed
to search for the centauro phenomenon at LHC.

It is important to note that  several scientists   working with 
emulsion chambers, and in particular  S.A.~Slavatinsky \cite{Slavatinsky}
argue that the data on primary particles  with an 
estimated energy   $E \aprge 10^{16}$~eV    exhibit  features
that  cannot be explained in the framework of the standard model.
One of these  features  is the existence of coplanar emission  \cite{Borisov}
(that is    families  where the  particles form approximately a
 straight line), another one is the indication 
of the existence of very strong Feynman scaling violations  in the
forward fragmentation region.
The   conclusion  that the  emulsion
chamber results in the energy range 10--100 PeV  indicate
new  unexpected effects  remain  very  controversial  and 
requires further analysis.

The emulsion chamber technique has now been in use for  over three decades,
and remains a  remarkable tool to study with  excellent resolution
the core  of high energy showers. The data obtained with these
detectors  has given
indications  of unusual phenomena, that  remain controversial  and unexplained.
An important open question  is the future of  these  studies, and how
to solve the puzzles they have  suggested.
Further progress requires   either a significant increase
in the   exposures  of the detectors (implying  larger  areas  and 
faster  analysis  methods),  or the introduction of new
innovative   experimental  methods  to study the  hadronic  core  of  showers.
An intersting new idea for an  hadronic  core detector  is  being developed
by the Tibet shower array \cite{Ding}.

\section{CERN: Opportunities and Challenges}

\subsection{Cosmic Ray Measurement at CERN}
At least three of the LEP experiments  at CERN have  taken data
on cosmic rays  not  only for the purpose of calibration,
but  in order  to  do  measurements.
The L3  detectors  \cite{L3} has  used the inner  detector  to measure
the inclusive muon  momentum spectrum  in the  range 15--2000~GeV.
This  measurement  is important to  constraint the calculation
of the atmospheric  neutrino fluxes in  a  similar range, since
both  $\mu$'s and  $\nu$'s  are produced in the decay of the same
primary mesons  ($\pi^\pm$'s and Kaons).
The L3 detector has also taken data in coincidence  with  a small
shower detector at the surface (the L3+C configuration) \cite{Wilkens}.
Multiple muon events  have also  been measured by the Aleph \cite{aleph} and
Delphi \cite{delphi} detectors.
Some  events are spectacular  containing  more than 100  parallel muons.
These events are  produced  by primary particles in the range
$10^{14}$--10$^{16}$~eV, and therefore  a detailed  study 
can provide  information about the spectrum and  composition
of cosmic rays in the knee region.  These  measurements are valuable
especially in combinations  with other measurements of showers  in the same
energy range, and in the spirit  of the ``bootstrap'' philosophy 
discussed above.

\subsection{LHC and cosmic ray physics}
The CERN  LHC project  (a 7+7 TeV $pp$ collider with options
for $p$--nucleus  and nucleus--nucleus  collisions)   has a   compelling
and ambitious program  that is clearly of  central  importance for
the development of fundamental science  \cite{Gianotti}. 
Four  detectors
(ATLAS, CMS, LHC-b and ALICE)
will  explore  the physics  at LHC,
LHC-b  is  dedicated to  the physics of the $b$--quark, ALICE to the study
of heavy--ion physics, 
ATLAS  and CMS  are   optimized  for the study of high $p_\perp$ interactions
between quark and  gluons, and the production of heavy particles
like  the Higgs or the  supersymmetric particles.

The essential contribution of  CERN to cosmic ray physics  is related
to the  more precise measurements of hadronic  interaction properties 
at high origin, and the LHC  project can  play a  fundamental  role.
The motivation is  clear:
in future studies of cosmic ray physics, very
high energy particles
will certainly have central importance, these particles
will be detected with indirect methods, 
 and the   precision and  resolution  of the measurements 
of the energy and   mass (or identity)  of the primary particles 
will depend on the   knowledge of hadronic interactions at a c.m.
energy as high as 400~TeV.
Measurements  that are {\em only}  possible  at LHC have the potential
to significantly improve the quality of these  measurements,
in the ``knee region'' and especially for the very high
energies   ($E \aprge 10^{19}$~eV)\footnote{An important task for
the c.r. community is to quantify more precisely  the  improvement
obtainable  with different  measurements at the LHC.} 

If the existence
of a  relatively large flux  of particles  beyond the GZK cutoff is  confirmed,
this  nearly certainly  implies the existence of ``new physics''
and the detailed  study of particles above the  expected  cutoff
will clearly
become one  of the most  important  fields of experimental studies
in fundamental  physics, 
however this  conclusion remains valid
 even in  the ``conservative'' scenario,
where   the  flux  exhibits the expected  cutoff and
all  cosmic  rays have a ``standard''  origin.
In this case the 
precise measurement of the  energy distribution  and composition
(together with the angular distribution) of particles 
at the ``end of the spectrum''
will be  essential
to obtain information about  the nature, location  and
time evolution of   the cosmic  accelerators  (that are
certainly going to be some  of the most  interesting  objects in the universe).

\subsection{Cross section measurements at CERN}
Of particular importance for cosmic  ray physics is the measurement of the
total and inelastic cross section. 
The Totem experiment  \cite{totem}    (designed  together with  CMS)
will provide measurements of the total cross section (with a precision of 1\%),
elastic scattering and diffractive processes at the LHC.
The total cross section will be measured using the luminosity
 independent method which based on the simultaneous
detection of elastic scattering at low momentum transfer and of 
the inelastic interactions.
One can use the  optical theorem:
\begin{equation}
\sigma_{\rm tot} = {4 \pi \over p_{\rm c.m.} } \; \Im {\rm m} [f(0)]
\end{equation}
where $f(\theta)$ is defined  by:
\begin{equation}
{d\sigma_{\rm el}  \over dt} = 
{\pi \over p_{\rm c.m.}^2 } \; 
{d\sigma_{\rm el}  \over d\Omega_{\rm c.m.} } =
{\pi \over p_{\rm c.m.}^2 } \; |f(\theta)|^2 
 \end{equation}
and the definition  of the total cross section:
\begin{equation}
{\cal } \, \sigma_{\rm tot} =
{\cal } \, (\sigma_{\rm el} +
\sigma_{\rm inel})  =
(N_{\rm el} +
N_{\rm inel})/{\cal L}  
\end{equation}
(where  ${\cal L}$ is the integrated luminosity)
to extract the  total  cross section independently  from the luminosity as:
\begin{equation}
\sigma_{\rm tot} = {16 \pi \over 1 + \rho^2} 
~ { [ d N_{\rm el} / dt  ]_{t = 0} \over (N_{\rm el} + N_{\rm inel}) } 
\end{equation}
where $\rho = \Re  {\rm e}[f(0)]/\Im  {\rm m}[f(0)] \simeq 0.15$ is  the ratio
of the real and imaginary part of the elastic scattering amplitude.
The difficulty of this   measurement is 
to  obtain
a good extrapolation  of the cross section for a
transfer momentum $t \to 0$.
Since  $-t = - (p_i - p_f)^2 = 2 p_{\rm c.m.}^2 (1-\cos \theta) \simeq
p_{\rm c.m.}^2 \, \theta^2$, this  implies the measurement at very small
angle.    The Totem experiment
 aims to a measurement down to a values
$-t \simeq 2 \times 10^{-2}$~GeV$^2$,  that corresponds to 
$\theta \simeq 20$~$\mu$rad, that   is  a displacement of 3 millimeters
at a distance of 150 meters from the interaction point.
The measurement is possible  with the use of the so called 
``Roman pots''\footnote{
The Roman pots are special devices mounted on the vacuum chamber of the 
accelerator. They  can be retracted to leave the 
vacuum chamber free for the  beam as required at the injection.
Once the final energy is attained and the circulating beams are stable, 
they can be moved close ($\sim 1$~mm) to the  beam.} 
 placed symmetrically on both sides of 
the intersection region  to detect protons
scattered at very small angles in elastic or quasi-elastic reactions.
A forward inelastic detector covering about 4 pseudorapidity 
units in the forward cones (from $\eta=3$ up to $\eta=7$)
with full azimuthal acceptance will be used to  measure
the  rate of inelastic reactions.
Totem has also  the potential  to measure diffractive  interactions.
As discussed before  a determination of $\sigma_{\rm diff}/\sigma_{\rm inel}$
is very important for shower development.

\subsection{Acceptance limitations and ``soft'' hadronic physics}
There are compelling   reasons  to expect that the most interesting
physics  at LHC 
will   involve  small cross
sections  (for example  the Higgs  production cross section is expected to
be  a  fraction of order $10^{-10}$ of  $\sigma_{\rm inel}$), 
and will manifest 
itself with  particle (jet)   production  at large $p_\perp$ and
therefore  at relatively  large  angles  with respect to the beam axis.
The  ATLAS and CMS  detectors  are  primarily designed to  study
this  type of processes, and their  acceptance  is limited  to the
angular  (pseudorapidity\footnote{The pseudorapidity
is defined  as $\eta = -\ln[\tan \theta/2]$.
 For  a massless particle it coincides with the  rapidity $y$  defined 
 differentially  as $dy = dp_\parallel/E$.})
 region $ |\eta| \le 2.5$ ($\theta \le 9.3^\circ$)
 for  charged particle  detection,
and to $|\eta| \le 5$ 
($\theta \le 0.77^\circ$) for calorimetric  energy flow.
\begin{figure}[htb]
\begin{center}
\epsfig{file=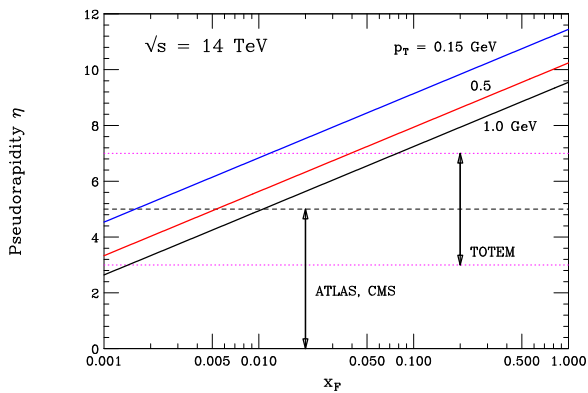,angle=90,width=7.5cm}
\caption{\small Angular (pseudorapidity) 
 regions that  corresponds to a  fixed $x_F$ 
at LHC.   The different curves correspond to three values of $p_\perp$.
The expected acceptance of three detectors are indicated.}
\label{fig:eta}
\end{center}
\end{figure}
As  discussed before the Totem detector will  cover 
the region $ 3 \le |\eta| \le 7$.
Fig.~\ref{fig:eta}   illustrates the  region of $x_F$ and $p_\perp$ that
 corresponds to these angular acceptances.  It is  clear that these detectors
will  give only poor results about particle  production  in the fragmentation 
regions  $|x_F| \aprge 0.1$.  We expect that  most of the energy of the
minimum---bias interactions ($> 90\%$)  will remain unobserved with the
detectors under construction.  

The design (and construction)
 of a full--acceptance detector for LHC is an extraordinarily 
 difficult  (and costly)  task .  A detailed project   for such a detector
 has been   elaborated by the Felix  collaboration \cite{felix}.
The scientific motivations for such a full (or very large) acceptance 
detectors are  also extensively discussed in \cite{felix}. 
While  there is  a consensus that the  scientific  priority
for the LHC science is in the large $p_\perp$ region,  there are strong
(even if less compelling)  arguments for 
a significant discovery potential exist  also for   detectors  that cover 
a larger  angular  region.  On the point of view of cosmic ray  studies
there is considerable interest in  having as  large
an acceptance as possible. 
The fragmentation region that plays a crucial role in 
shower development corresponds to the
pseudorapidity range $ 6 \aprle |\eta| \aprle 10$.
The possibility to explore, even partially,  a broader  phase space
region at LHC upgrading  the approved detectors
 certainly deserves to be investigated energetically.

\section{Conclusions}
The  next decade looks very interesting for cosmic ray studies,
in fact it is possible that finally an understanding  of the  main sources
of the high energy radiation in the universe  will be obtained.
This  understanding will   be the result of a large  experimental effort,
including  direct and  indirect measurements of cosmic rays.
For a correct interpretation  of the EAS  shower measurements
the contribution of  accelerator  experiments  will be of great importance,
in the entire  energy range 
between $10^{15}$ and 10$^{20}$~eV.
The desired  measurements at LHC are not easy to perform, and an optimum
program would  require a larger   acceptance coverage and therefore
additional costly instrumentation; it can however be argued that these
measurements are not only of relevance  for cosmic ray research  but have
also a  significant intrinsic interest and   the potential for
scientific discovery, and deserve careful  analysis.

\vspace{0.35 cm}
\noindent {\bf Acknowledgments}  I'm grateful to many people
for discussions and clarifications, in particular to
A. Castellina,
K.~Eggert,
J.~Ellis,
R.~Engel,
T.~Gaisser,
K.H.~Kampert,
J.~Kempa,
J.~Knapp,
A.~Ohsawa,
B.~Pattison,
L.~Resvanis,
O.~Saavedra,
G.~Schatz,
E.~Shibuya,
S.~Slavatinsky
T.~Stanev,
and S.~Vernetto.

\end{document}